\documentstyle[manuscript,aps]{revtex}
\begin{document}
\preprint{
\begin{tabular}{r}
MIT-CTP-2673 
\end{tabular}
} 
\begin{titlepage}
\title{Two Component Theory of Neutrino Flavor Mixing}
\author{Elisabetta Sassaroli \thanks{This work is supported in part by funds 
provided by the U.S. Department of Energy (D.~O.~E.) under 
cooperative research agreement $\#$DF-FC02-94ER40818 and in 
part by the University and INFN of Perugia.}  }
\address{Laboratory for Nuclear Science and Department of Physics}
\address{Massachusetts Institute of Technology}
\address{Cambridge, MA 02139}
\address{Dipartimento di Fisica and sezione INFN} 
\address{Universita\' \  di Perugia} 
\address {I-06123 Perugia, Italy}
\maketitle
\begin{abstract}
Neutrino flavor  mixing is discussed in terms of two-component coupled
left-handed flavor fields. This is to take into account the fact that the 
weak interaction couples only to left-handed fields. The flavor 
fields are written 
through a rotation matrix, as a linear combination  of 
left-handed free fields.  
In order to obtain properly normalized wave functions directly from those 
free fields, states of mixed helicity have to be considered.
Neutrino flavor oscillation amplitudes are also derived.
 
\end {abstract}

\thispagestyle{empty}

\end{titlepage}

\section{Introduction}

In Ref. \cite{elisa} we have discussed neutrino flavor mixing by assuming
that neutrinos are described by Dirac fields. However to take into account
the fact that neutrinos are created with negative chiralities, we were
forced to consider only left-handed wave functions as observable wave
functions.

In this paper we want to consider a Lagrangian which is only a function of
left-handed, but not right-handed fields.

More specifically, neutrino flavor mixing is discussed in terms of the
two-component fields $\psi _{e_L}$ and $\psi _{\mu _L},$ defined by 
\begin{equation}
\psi _{_{eL}}=\frac 12(1-\gamma _5)\psi _e,\ \ \ \psi _{_{\mu L}}= \frac 12%
(1-\gamma _5)\psi _\mu , 
\end{equation}
with $\psi _e$ and $\psi _\mu $ Dirac fields. These fields satisfy the
following Lagrangian \cite{bil} 
\begin{equation}
\begin{array}{c}
{\cal L}={\cal L}_e+{\cal L}_\mu +{\cal L}_{int}, \\ {\cal L}_e= i{\bar \psi 
}_{eL}\gamma \cdot \partial \psi _{eL}-\frac{m_e}2({\bar \psi }_{eL}\psi
_{eL}^c+{\bar \psi }_{eL}^c\psi _{eL}^{}), \\ {\cal L}_\mu =i{\bar \psi }%
_{\mu L}\gamma \cdot \partial \psi _{_\mu L}-\frac{m_\mu }2({\bar \psi }%
_{\mu L}\psi _{_\mu L}^c+{\bar \psi }_{\mu L}^c\psi _{\mu L}^{}), \\ {\cal L}%
_{int}=-\frac \delta 2[{\bar \psi }_{eL}\psi _{\mu L}^c+{\bar \psi }_{\mu
L}^c\psi _{eL}^{}+{\bar \psi }_{\mu L}\psi _{eL}^c+{\bar \psi }_{eL}^c\psi
_{_\mu L}^{}], 
\end{array}
\end{equation}
where $m_e$ and $m_\mu $ are the electron and muon neutrino masses
respectively, $\delta $ is the coupling constant, and $\psi _{eL}^c$ and $%
\psi _{\mu L}^c$ are the so called charge-conjugate fields defined in the
chiral representation by 
\begin{equation}
\psi _{_{eL}}^c=-i\gamma ^2(\psi _{eL}^{\dagger })^T,\ \ \ \psi _{_{\mu
L}}^c=-i\gamma ^2(\psi _{\mu L}^{\dagger })^T. 
\end{equation}

We assume that neutrinos are produced through some weak interaction process
and then they mix in flavor accordingly to the Lagrangian given by Eq. (2).

The Lagrangian given above allows for exact diagonalization. The rotation
matrix between the coupled fields $\psi _{eL}$ and $\psi _{\mu L}$ and the
free uncoupled fields $\psi _{1L}$ and $\psi _{2L},$ which diagonalizes Eq.
(2), is the same as the one obtained in Ref .\cite{elisa}.

The neutrino left-handed fields $\psi _{eL}$ and $\psi _{\mu L}$ are written
as linear combinations of the left-handed free fields $\psi _{1L}$ and $\psi
_{2L}$ through this rotation matrix. The two-component free fields $\psi
_{1,2L}$ are described by the Lagrangian of the Majorana type \cite{case} 
\begin{equation}
{\cal L}_{}=i{\bar \psi }_L\gamma \cdot \partial \psi _L-\frac{m_{}}2({\bar 
\psi }_L\psi _L^c+{\bar \psi }_L^c\psi _L^{}). 
\end{equation}

Therefore there is no distinction in this theory between particles and
anti-particles.

Neutrino flavor wave functions are derived as matrix elements of the fields $%
\psi _{1L}$ and $\psi _{2L}$ between the vacuum state and one-particle
states. In order to obtain properly normalized wave functions directly from
those fields, a suitable combination of one particle states has to be
considered, as described in the following section.

Neutrino flavor oscillation probability amplitudes are also derived. The
standard neutrino oscillation probabilities, derived in the literature from
a quantum mechanical treatment, are recovered in the field theory treatment
only in the relativistic limit. The same limitation occurs also for Dirac
fields.

The paper is organized as follows. In Sec. 2, the lagrangian density, given
by Eq. (4) is considered. In Sec. 3, the Lagrangian (2) is diagonalized and
neutrino flavor oscillations are discussed. The last section closes with
some concluding remarks.

\section{Left-Handed (Majorana) Fermions}

As stated in the Introduction, we will review and study here the following
Lagrangian 
\begin{equation}
{\cal L}_{}=i{\bar \psi }_L\gamma \cdot \partial \psi _L-\frac{m_{}}2({\bar 
\psi }_L\psi _L^c+{\bar \psi }_L^c\psi _L^{}), 
\end{equation}
which differs by a total derivative from the Majorana lagrangian density
(see for example \cite{boe,kay,bah,pal,kim}).

In the chiral representation the left-handed fields $\psi _L$ and $\psi _L^c$
are 
\begin{equation}
\psi _{_L}=\frac 12(1-\gamma _5)\psi =\left( 
\begin{array}{c}
\phi \\ 
0 
\end{array}
\right) ,\ \ \ \ \psi _L^c=-i\gamma ^2\psi ^{\dagger T}=\left( 
\begin{array}{c}
0 \\ 
i\sigma ^2\phi ^{\dagger T} 
\end{array}
\right) 
\end{equation}
where $\psi =\left( 
\begin{array}{c}
\phi \\ 
\chi 
\end{array}
\right) ,$ $\phi $ and $\chi $ are two-component fields, in particular 
\begin{equation}
\phi =\left( 
\begin{array}{c}
\phi _1 \\ 
\phi _2 
\end{array}
\right) ,\ \ \ \ \ \ \ \ \phi ^{\dagger T}=\left( 
\begin{array}{c}
\phi _1^{*} \\ 
\phi _2^{*} 
\end{array}
\right) . 
\end{equation}
In terms of the fields $\phi $ and $\phi ^{\dagger }$, the Lagrangian given
by Eq. (5) becomes 
\begin{equation}
{\cal L}_{}=i{\phi }^{\dagger }\overline{\sigma }\cdot \partial \phi +\frac{%
im}2(\phi ^T\sigma ^2\phi -\phi ^{\dagger }\sigma ^2\phi ^{\dagger T}), 
\end{equation}
with $\overline{\sigma }=(1,-\overrightarrow{\sigma }).$

Because the matrix $\sigma ^2$ is anti-symmetric the mass terms $\phi
_\alpha \sigma _{\alpha \beta }^2\phi _\beta $ and $\phi _a^{\dagger }\sigma
_{\alpha \beta }^2\phi _\beta ^{\dagger }$ are identically zero. Therefore
the Lagrangian given by Eq. (8) makes sense only if the fields $\phi $ and $%
\phi ^{\dagger }$ are considered as Grassmann fields, i.e. their components
satisfy the conditions 
\begin{equation}
\phi _1\phi _2+\phi _2\phi _1=0,\ \ \ \ \phi _1^{*}\phi _2^{*}+\phi
_2^{*}\phi _1^{*}=0. 
\end{equation}
The equation of motion for the two-component field $\phi $ is 
\begin{equation}
i\overline{\sigma }\cdot \partial \phi -im\sigma ^2\phi ^{\dagger T}=0. 
\end{equation}
We first solve Eq. (10) for the momentum $p$ along the z-axis and then we
generalize the result to three dimensions. We consider the ansatz 
\begin{equation}
\phi _p(z,t)=a_pu_pe^{-iEt}e^{ipz}+a_p^{\dagger }v_pe^{iEt}e^{-ipz}, 
\end{equation}
where $a_p$ and $a_p^{\dagger }$ are Grassmann numbers and $u$ and $v$ are
two component c-number spinors 
\begin{equation}
u_p=\left( 
\begin{array}{c}
u_1 \\ 
u_2 
\end{array}
\right) ,\ \ \ \ v_p=\left( 
\begin{array}{c}
v_1 \\ 
v_2 
\end{array}
\right) . 
\end{equation}
By substituting Eq. (11) into Eq. (10) we obtain two set of solutions, one
is 
\begin{equation}
u_1=\frac m{E+p}v_2^{*},\ \ \ \ v_1^{}=\frac{-m}{E+p}u_2^{*}, 
\end{equation}
and the other one is 
\begin{equation}
u_2=\frac{-m}{E-p}v_1^{*},\ \ \ \ v_2^{}=\frac m{E-p}u_1^{*}. 
\end{equation}
The solutions are equivalent if 
\begin{equation}
E=\pm \sqrt{p^2+m^2}. 
\end{equation}

The solution to Eq. (10) for a given momentum p and positive energy $E=\sqrt{%
p^2+m^2}$ can be written as 
\begin{equation}
\begin{array}{c}
\phi _p(z,t)=a(p,1)_{}v_2^{*}\frac m{E+p}\left( 
\begin{array}{c}
1 \\ 
0 
\end{array}
\right) e^{-iEt}e^{ipz}+a_{}^{\dagger }(p,1)v_2\left( 
\begin{array}{c}
0 \\ 
1 
\end{array}
\right) e^{iEt}e^{-ipz} \\ 
+a(p,2)u_2\left( 
\begin{array}{c}
0 \\ 
1 
\end{array}
\right) e^{-iEt}e^{ipz^{}}-a_{}^{\dagger }(p,2)\frac m{E+p}u_2^{*}\left( 
\begin{array}{c}
1 \\ 
0 
\end{array}
\right) e^{iEt}e^{-ipz}. 
\end{array}
\end{equation}

The solution corresponding to $E=-\sqrt{p^2+m^2}$ is equivalent to the one
above with the substitution $p\rightarrow -p^{\prime },$ therefore in this
theory we have only one type of particle.

It is easy to see that the components of $\phi _p(z,t)$ satisfies the
condition, given by Eq. (9), if the operators $a(p,1)$ , $a(p,2)$ satisfy
the Grassman algebra 
\begin{equation}
\{a(p,i)_{,}a_{}^{\dagger }(p,j)\}=\delta _{ij},\ \ \ 
\end{equation}
and 
\begin{equation}
|u_2|=|v_2|. 
\end{equation}

The phase between $u_2,$ $v_2$ can be chosen in such a way that $%
|u_2|=|v_2|=\lambda _p,$ where $\lambda _p$ is a normalization constant.

The general solution is given therefore by 
\begin{equation}
\phi (z,t)=\frac 1{\sqrt{L}}\sum_p\lambda _p\left[ 
\begin{array}{c}
( 
\frac m{E+p}a(p,1)\chi ^{(1)}+a(p,2)\chi ^{(2)})e^{-iEt}e^{ipz} \\ +\left(
a_{}^{\dagger }(p,1)\chi ^{(2)}-\frac m{E+p}a_{}^{\dagger }(p,2)\chi
^{(1)}\right) e^{iEt}e^{-ipz} 
\end{array}
\right] , 
\end{equation}

with 
\begin{equation}
\chi ^{(1)}=\left( 
\begin{array}{c}
1 \\ 
0 
\end{array}
\right) ,\ \ \ \chi ^{(2)}=\left( 
\begin{array}{c}
0 \\ 
1 
\end{array}
\right) 
\end{equation}

The result given by Eq. (19) can be generalized in three spatial dimensions
as follows. Given the spin base $\chi ^{(1)},$ $\chi ^{(2)},$ such that 
\begin{equation}
\overrightarrow{\sigma \cdot }\overrightarrow{p}\chi ^{(1)}=p\chi ^{(1)},\ \
\ \ \overrightarrow{\sigma \cdot }\overrightarrow{p}\chi ^{(2)}=-p\chi
^{(2)}, 
\end{equation}
with $p=|\overrightarrow{p}|,$ for example 
\begin{equation}
\chi ^{(1)}=\left( 
\begin{array}{c}
e^{- 
\frac i2}\cos (\frac \theta 2) \\ e^{+\frac i2}\sin (\frac \theta 2) 
\end{array}
\right) ,\ \ \ \ \chi ^{(2)}=\left( 
\begin{array}{c}
-e^{- 
\frac i2}\sin (\frac \theta 2) \\ e^{+\frac i2}\cos (\frac \theta 2) 
\end{array}
\right) , 
\end{equation}
we can write the two spinors $u_p$ and $v_p$ defined in Eq. (12) as a linear
combination of the spin base defined in Eq. (21). The ansatz, given in Eq.
(11), can be therefore generalized as 
\begin{equation}
\phi _{{\bf p}}({\bf x},t)=a_{{\bf p}}(u_1\chi ^{(1)}+u_2\chi
^{(2)})e^{-iEt}e^{i{\bf p\cdot x}}+a_{{\bf p}}^{\dagger }(v_1\chi
^{(1)}+v_2\chi ^{(2)})e^{iEt}e^{-i{\bf p\cdot x}}. 
\end{equation}
By inserting Eq. (23) into the equation of motion (10) and by noticing that 
\begin{equation}
i\sigma ^2\chi ^{*(1)}=-\chi ^{(2)},\ \ \ \ i\sigma ^2\chi ^{*(2)}=\chi
^{(1)}, 
\end{equation}
we get the same solutions as given in Eqs. (13), (14), (15), (17) and (18).

Therefore the general solution can be written as 
\begin{equation}
\phi ({\bf x},t)=\frac 1{\sqrt{V}}\sum_{{\bf p}}\lambda _p\left[ 
\begin{array}{c}
\left( 
\frac m{E+p}a({\bf p,}1{\bf )}\chi ^{(1)}+a({\bf p,}2{\bf )}\chi
^{(2)}\right) e^{-ip\cdot x} \\ +\left( a_{}^{\dagger }({\bf p,}1{\bf )}\chi
^{(2)}-\frac m{E+p}a_{}^{\dagger }({\bf p,}2{\bf )}\chi ^{(1)}\right)
e^{ip\cdot x} 
\end{array}
\right] , 
\end{equation}

In terms of its components, the field $\phi ({\bf x},t)$ can be written as 
\begin{equation}
\begin{array}{c}
\phi^{(1)}( 
{\bf x},t)=\frac 1{\sqrt{V}}\sum_{{\bf p}}\lambda _p\frac m{E+p}\left[ a(%
{\bf p,}1{\bf )}e^{-ip\cdot x}-a_{}^{\dagger }({\bf p,}2{\bf )}e^{ip\cdot
x}\right] , \\ 
\begin{array}{c}
\phi^{(2)}( 
{\bf x},t)=\frac 1{\sqrt{V}}\sum_{{\bf p}}\lambda _p\left[ a({\bf p}%
,2)_{}e^{-ip\cdot x}+a_{}^{\dagger }({\bf p,}1)e^{ip\cdot x}\right] . \\  
\end{array}
\end{array}
\end{equation}

The constant $\lambda _p$ is chosen in such a way that the equal time
anti-commutation relations ($i,j=1,2)$%
\begin{equation}
\{\phi ^{(i)}({\bf x},t),\phi ^{(j)\dagger }({\bf x}^{\prime },t)\}=\delta
_{ij}\delta ^{(3)}({\bf x}-{\bf x}^{\prime }),\ \ \ \ \{\phi ^{(i)}({\bf x}%
,t),\phi ^{(j)}({\bf x}^{\prime },t)\}=0,\ \ \ \{\phi ^{(i)\dagger }({\bf x,}%
t),\phi ^{(j)\dagger }({\bf x,}t)\}=0,
\end{equation}
hold, i.e. 
\begin{equation}
\lambda _p=\sqrt{\frac{E+p}{2E}}.
\end{equation}
The field $\phi ({\bf x},t)$ describes a particle which can be in two
different states of helicity, corresponding to the states $\chi ^{(1)}$ and $%
\chi ^{(2)}.$ However the physical interpretation of the field $\phi ({\bf x}%
,t)$ poses some problems. For example, the one-particle state $|{\bf p}%
,i>=a_{{\bf p}i}^{\dagger }|0>$ gives wave functions which are not properly
normalized 
\begin{equation}
\begin{array}{c}
<0|\phi (
{\bf x},t)|{\bf p,}1>=\frac 1{\sqrt{V}}\lambda _p\frac m{E+p}\chi
^{(1)}e^{-ip\cdot x}, \\ <0|\phi ({\bf x},t)|{\bf p},2>=\frac 1{\sqrt{V}}%
\lambda _p\chi ^{(2)}e^{-ip\cdot x}.
\end{array}
\end{equation}
The only way to obtain properly normalized wave functions is to consider
states such as 
\begin{equation}
|\phi >_{pL}=|{\bf p},1>+|{\bf p},2>.
\end{equation}
The wave function associated with $|\phi >_{pL}$is 
\begin{equation}
\phi ({\bf x,}t)_L=\frac 1{\sqrt{V}}\lambda _p[\frac m{E+p}\chi ^{(1)}+\chi
^{(2)}]e^{-ip\cdot x}.
\end{equation}
This wave function describes a left-handed particle which is in a state of
mixed helicity with the positive helicity suppressed by the factor $\frac m{%
E+p}.$

\section{Flavor Mixing}

The Lagrangian given by Eq. (2) can be written in terms of the fields $\phi
_e$ and $\phi _\mu $ where 
\begin{equation}
\psi _{eL}=\left( 
\begin{array}{c}
\phi _e \\ 
0 
\end{array}
\right) ,\ \ \ \psi _{\mu L}=\left( 
\begin{array}{c}
\phi _\mu \\ 
0 
\end{array}
\right) 
\end{equation}
as 
\begin{equation}
\begin{array}{c}
{\cal L}=i\overline{\phi }_e\overline{\sigma }\cdot \partial \phi _e+\frac{%
im_e}2(\phi _e^T\sigma ^2\phi _e^{}-{\phi }_e^{\dagger }\sigma ^2\phi
_e^{T\dagger }) \\ +i 
\overline{\phi }_\mu \overline{\sigma }\cdot \partial \phi _\mu +\frac{%
im_\mu }2(\phi _\mu ^T\sigma ^2\phi _\mu ^{}-{\phi }_\mu ^{\dagger }\sigma
^2\phi _\mu ^{T\dagger }) \\ +i\frac \delta 2\left[ \phi _e^T\sigma ^2\phi
_\mu ^{}-{\phi }_e^{\dagger }\sigma ^2\phi _\mu ^{T\dagger }+\phi _\mu
^T\sigma ^2\phi _e-{\phi }_\mu ^{\dagger }\sigma ^2\phi _e^{T\dagger
}\right] . 
\end{array}
\end{equation}
It is possible to see that the rotation matrix U, defined in \cite{elisa} as 
\begin{equation}
U=\left( 
\begin{array}{cc}
\frac 1{\sqrt{1+M_1^2}} & \frac{M_1}{\sqrt{1+M_1^2}} \\ \frac{M_1}{\sqrt{%
1+M_1^2}} & \frac{-1}{\sqrt{1+M_1^2}} 
\end{array}
\right) , 
\end{equation}
with $M_1$ given by 
\begin{equation}
M_1=\frac{m_\mu -m_e+R}{2\delta },\ \ \ \ R=\sqrt{(m_\mu -m_e)^2+4\delta ^2} 
\end{equation}
when applied to the fields $\phi _1$ and $\phi _2$%
\begin{equation}
\phi _\nu =\left( 
\begin{array}{c}
\phi _e \\ 
\phi _\mu 
\end{array}
\right) =U\left( 
\begin{array}{c}
\phi _1 \\ 
\phi _2 
\end{array}
\right) , 
\end{equation}
uncouples the Lagrangian given by Eq. (33). The uncoupled Lagrangian is
given by

\begin{equation}
\begin{array}{c}
{\cal L}_D=[i\overline{\phi }_1\overline{\sigma }\cdot \partial \phi _1+%
\frac{im_1}2(\phi _1^T\sigma ^2\phi _1-\phi _1^{\dagger }\sigma ^2\phi
_1^{\dagger T})] \\ + \lbrack i\overline{\phi }_2\overline{\sigma }\cdot
\partial \phi _2+\frac{im_2}2(\phi _2^T\sigma ^2\phi _2-\phi _2^{\dagger
}\sigma ^2\phi _2^{\dagger T}) 
\end{array}
\end{equation}
with 
\begin{equation}
m_{1,2}={\frac 12}[(m_e+m_\mu )\pm R]. 
\end{equation}

The fields $\phi _1$ and $\phi _2$ which have been discussed in Se. 2, are
free fields of masses $m_1$, $m_2$ respectively 
\begin{equation}
\phi _1({\bf x},t)={\frac 1{\sqrt{V}}}\sum_{{\bf p}}\lambda _{1p}\left[ 
\begin{array}{c}
\left( 
\frac m{E_1+p}a_1({\bf p,}1)\chi ^{(1)}+a_1({\bf p,}2)\chi ^{(2)}\right) {%
e^{-iE_1t}e^{i{\bf p}\cdot {\bf x}}} \\ + \left( a_1^{\dagger }({\bf p,}%
1)\chi ^{(2)}-\frac m{E_1+p}a_1^{\dagger }({\bf p,}2)\chi ^{(1)}\right) {%
e^{iE_1t}e^{-i{\bf p}\cdot {\bf x}}} 
\end{array}
\right] , 
\end{equation}
\begin{equation}
\phi _2({\bf x},t)={\frac 1{\sqrt{V}}}\sum_{{\bf p}}\lambda _{2p}\left[ 
\begin{array}{c}
\left( 
\frac m{E_2+p}a_2({\bf p,}1)\chi ^{(1)}+a_2({\bf p,}2)\chi ^{(2)}\right) {%
e^{-iE_2t}e^{i{\bf p}\cdot {\bf x}}} \\ + \left( a_2^{\dagger }({\bf p,}%
1)\chi ^{(2)}-\frac m{E+p}a_2^{\dagger }({\bf p,}2)\chi ^{(1)}\right) {e}%
^{-iE_2t}{e^{-i{\bf p}\cdot {\bf x}}}_{} 
\end{array}
\right] , 
\end{equation}
where 
\begin{equation}
E_1=\sqrt{p^2+m_1^2},\ \ \ \ E_2=\sqrt{p^2+m_2^2}, 
\end{equation}
\begin{equation}
\lambda _{1p}=\sqrt{\frac{E_1+p}{2E_1}},\ \ \ \ \lambda _{2p}=\sqrt{\frac{%
E_2+p}{2E_2}} 
\end{equation}

The electron and neutrino field operators ${\phi }_e$ and ${\phi }_\mu $ are
related to the diagonal (uncoupled) field operators ${\phi }_1$ and ${\phi }%
_2$ through the rotation matrix $U$ defined in Eq. (34).

For a given momentum ${\bf p}$ and spin $i=1,2,$ there are two possible
one-particle states, one associated with the field ${\phi }_1$ and the other
one with the field ${\phi }_2$, i.e. 
\begin{equation}
a_1^{\dagger }({\bf p},i)|0>=|1({\bf p},i)>,\ \ \ \ a_2^{\dagger }({\bf p,}%
i)|0>=|2({\bf p},i)>. 
\end{equation}

The wave function associated with the state 
\begin{equation}
|\phi _1>_{pL}=|1({\bf p},1)>+|1({\bf p},2)>, 
\end{equation}
is 
\begin{equation}
\begin{array}{c}
\phi _{L\nu }({\bf x},t)=\left( 
\begin{array}{c}
\phi _{Le}( 
{\bf x},t) \\ \phi _{L\mu }({\bf x},t) 
\end{array}
\right) =\left( 
\begin{array}{c}
<0|\phi _e( 
{\bf x},t)|\phi _1>_{pL} \\ <0|\phi _\mu ({\bf x},t)|\phi _1>_{pL} 
\end{array}
\right) = \\ 
=\left( 
\begin{array}{c}
\frac 1{\sqrt{1+M_1^2}} \\ \frac{M_1}{\sqrt{1+M_1^2}} 
\end{array}
\right) \frac 1{\sqrt{V}}\lambda _{_1p}[\frac m{E_1+p}\chi ^{(1)}+\chi
^{(2)}]e^{i{\bf p}\cdot {\bf x}}e^{-iE_1t}{}^{}. 
\end{array}
\end{equation}

This represents a plane wave of mixed helicity. In any location inside the
volume V there is a probability equal to $({\frac 1{1+M_1^2}})$ of finding
the neutrino in the electron flavor and probability equal to $({\frac{M_1^2}{%
1+M_1^2}})$ of finding it in the muon flavor.

Similar consideration can be applied for the other state of given momentum $%
|\phi _2>_{pL}=|2({\bf p},1)>+|2({\bf p},2)>.$

To be able to describe neutrino flavor oscillations, we need to consider a
linear combinations of the states $|\phi _1>_{pL}$ and $|\phi _2>_{pL},$
such as 
\begin{equation}
|\phi >_L=A|\phi _1>_{pL}+B|\phi _2>_{pL}, 
\end{equation}
with 
\begin{equation}
|A|^2+|B|^2=1. 
\end{equation}

The matrix element 
\begin{equation}
<0|\phi _e({\bf x},t)|\phi >_L=\phi _{eL}({\bf x},t)={\frac 1{\sqrt{V}}}{%
\frac 1{\sqrt{1+M_1^2}}}\left[ 
\begin{array}{c}
A\lambda _{_1p}[ 
\frac m{E_1+p}\chi ^{(1)}+\chi ^{(2)}]e^{-iE_1t} \\ +M_1B\lambda _{_2p}[%
\frac m{E_2+p}\chi ^{(1)}+\chi ^{(2)}]e^{-iE_2t} 
\end{array}
\right] e^{i{\bf p}\cdot {\bf x}}, 
\end{equation}
gives the probability amplitude of finding a neutrino of momentum ${\bf p}$
at the space-time point $({\bf x},t)$ with the electron flavor. In the same
way, the matrix element 
\begin{equation}
<0|\phi _\mu ({\bf x},t)|\phi >_L=\phi _{\mu L}({\bf x},t)={\frac 1{\sqrt{V}}%
}{\frac 1{\sqrt{1+M_1^2}}}\left[ 
\begin{array}{c}
M_1A\lambda _{_1p}[ 
\frac m{E_1+p}\chi ^{(1)}+\chi ^{(2)}]e^{-iE_1t} \\ - B\lambda _{_2p}[\frac m%
{E_2+p}\chi ^{(1)}+\chi ^{(2)}]e^{-iE_2t} 
\end{array}
\right] e^{i{\bf p}\cdot {\bf x}}, 
\end{equation}
is the probability amplitude for the muon flavor.

The coefficients $A$ and $B$ in Eqs. (48) and (49) are determined through
the initial boundary conditions.

However, it is possible only in the relativistic limit, when the term $\frac 
m{E+p}$ $\simeq 0$ in Eqs. (48), (49), to have only one given flavor. This
limitation occurs also for Dirac fields. The procedure illustrated in \cite
{elisa} to obtain neutrino flavor oscillation amplitudes is valid in the
relativistic limit, because the condition given by Eq. (47) holds only in
that limit. Other authors by using different approaches have found the same
limition. \cite{giu,bla}

In the relativistic limit, Eqs. (48) (49) can be approximated as 
\begin{equation}
\phi _{eL}({\bf x},t)\simeq {\frac 1{\sqrt{V}}}{\frac{e^{i{\bf p}\cdot {\bf x%
}}}{\sqrt{1+M_1^2}}}(Ae^{-iE_1t}+M_1Be^{-iE_2t})\chi ^{(2)}, 
\end{equation}
\begin{equation}
\phi _{\mu L}({\bf x},t)\simeq {\frac 1{\sqrt{V}}}{\frac{e^{i{\bf p}\cdot 
{\bf x}}}{\sqrt{1+M_1^2}}(A}M_1e^{-iE_1t}-Be^{-iE_2t})\chi ^{(2)}. 
\end{equation}
The coefficients $A$ and $B$ are determined from the initial boundary
condition. Suppose for example that at $t=0$%
\begin{equation}
\phi _{\mu L}({\bf x},t=0)=0, 
\end{equation}
so we have only the electron flavor present. The other one is obtained by
the normalization condition 
\begin{equation}
\int_V{d^3{\bf x}|\psi _{eL}({\bf x},t=0)|^2}=1. 
\end{equation}

By imposing the boundary conditions given by Eq. (52) and Eq. (53) we obtain
the following flavor wave functions 
\begin{equation}
\phi _{eL}({\bf x},t)={\frac{e^{i{\bf p}\cdot {\bf x}}}{\sqrt{V}}}{\frac 1{%
1+M_1^2}}[e^{-iE_1t}+M_1^2e^{-iE_2t}]\chi ^{(2)}, 
\end{equation}
\begin{equation}
\phi _{{\mu }L}({\bf x},t)={\frac{e^{i{\bf p}\cdot {\bf x}}}{\sqrt{V}}}{%
\frac{M_1}{1+M_1^2}}[e^{-iE_1t}-e^{-iE_2t}]\chi ^{(2)}. 
\end{equation}
These amplitudes squared give the standard neutrino oscillation
probabilities as described in \cite{elisa}.

\section{Conclusions}

We have discussed a model to describe neutrino flavor mixing which takes
into account the fact that neutrinos are created through weak interaction,
which couples only to left-handed fields. The flavor wave functions are in a
superposition of states of mixed helicities. The standard neutrino
oscillation probabilities are obtained in the relativistic limit.

\end{document}